\documentclass[%
 aps,
 prl,%
 amsmath,amssymb,
 reprint,
]{revtex4-1}

\draft % marks overfull lines with a black rule on the right

\usepackage{graphicx}% Include figure files
\usepackage{dcolumn}% Align table columns on decimal point
\usepackage{bm}% bold math
\usepackage{color}

\begin{document}

% Use the \preprint command to place your local institutional report number 

% on the title page in preprint mode.

% Multiple \preprint commands are allowed.

%\preprint{}

\title{Acoustically tuneable optical transmission \\ through a subwavelength hole with a bubble}
\author{Ivan S. Maksymov and Andrew D. Greentree}
%\email{ivan.maksymov@rmit.edu.au} %% email address is required
\affiliation{ARC Centre of Excellence for Nanoscale BioPhotonics, School of Science, RMIT University, Melbourne, VIC 3001, Australia}

\date{\today}

\begin{abstract}

{\color{black} Efficient manipulation of light with sound in subwavelength-sized volumes is important for applications in photonics, phononics and biophysics, but remains elusive. We theoretically demonstrate the control of light with MHz-range ultrasound in a subwavelength, $300$~nm wide water-filled hole with a $100$~nm radius air bubble. Ultrasound-driven pulsations of the bubble modulate the effective refractive index of the hole aperture, which gives rise to spectral tuning of light transmission through the hole. This control mechanism opens up novel opportunities for tuneable acousto-optic and optomechanical metamaterials, and all-optical ultrasound transduction.}

\end{abstract}

%\pacs{}

\maketitle %\maketitle must follow title, authors, abstract and \pacs

The presence of a single circular hole with the diameter $w$ in an opaque metal film, with $w$ much smaller than the wavelength of incident light $\lambda_{\rm{0}}$, leads to optical phenomena unpredicted by the classical aperture theories \cite{Gen07, Aba07, thesis}. Such phenomena include enhanced transmission of light through the hole \cite{Gen07, Aba07, thesis}, Purcell effect and emission of nonclassical light \cite{Bul11, Mak10}, as well as spectroscopy and sensing \cite{Gor08}. These functionalities are achievable in both single and periodically arrayed holes, and they are due to the interaction of light with surface plasmon (SP) resonances at the surface of the metal film and Fabry-Perot resonances of guided optical modes inside the hole \cite{Gen07, Aba07, thesis}.

The Fabry-Perot resonances give rise to a peak in light transmission through a single hole when $\lambda_{\rm{0}} \approx \lambda_{\rm{c}}$, where $\lambda_{\rm{c}} \propto w n_{\rm{f}}$ is the cutoff wavelength of the fundamental guided mode of the hole and $n_{\rm{f}}$ the refractive index of the material filling the hole \cite{Mak10, thesis}. Thus, the transmission peak becomes spectrally tuneable by either changing $w$ or controlling $n_{\rm{f}}$ \cite{Aba07, thesis}. This has been achieved by using stretchable metal nanovoids \cite{Col09}, liquid crystals \cite{Liu11}, nonlinear optical materials \cite{Smo02}, and electrically tuneable semiconductor materials \cite{Sha07}. Spectral tuning of transmission has also been demonstrated by applying external magnetic fields \cite{Mak16_review} and using surface acoustic waves \cite{Ger07}.

{\color{black} Other types of electromagnetic waves, e.g. microwaves, may also be transmitted through subwavelength apertures in a fashion similar to light \cite{Mak14}. Enhanced transmission of sound through acoustically subwavelength apertures has also been observed \cite{Chr07, Est08}.

The coexistence of the waves of different nature in the same structure allows controlling one wave with another (e.g., light in holes filled with a magneto-dielectric material may be controlled with microwaves and vice-versa \cite{Mak15}). However, despite the progress in the control of the interaction of light with structural deformations of micro- and nanostructures \cite{Asp14, phoxonic}, spectral tuning of the optical transmission through subwavelength apertures with ultrasound remains elusive.}

\begin{figure}[tb]
\centering\includegraphics[width=8.5cm]{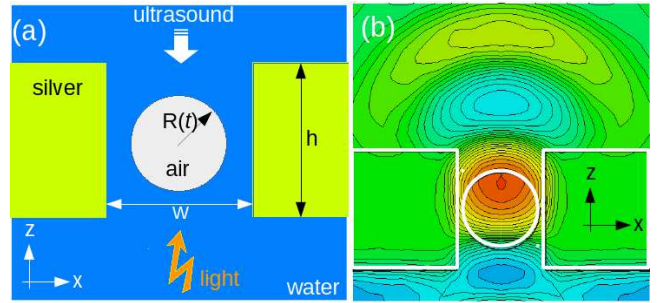}
\caption{(a) Schematic cross-section of the water-filled hole with an air bubble. The thickness of the silver thin film is $h=400$~nm. The diameter of the hole is $w=300$~nm. The wavelength of ultrasound is two orders of magnitude larger than $w$. (b) Instantaneous $E_{\rm{x}}$ electric field amplitude snapshot in the cross-section of the hole with a bubble. The incident optical plane wave propagates in the $+z$-direction. False colours: red -- maximum, green -- zero, and blue -- minimum.}
\label{fig1}
\end{figure}

{\color{black} In this Letter, we theoretically demonstrate a spectral tuning of light transmission with sound in a subwavelength-sized volume. A water-filled, $300$~nm wide circular hole in a $400$~nm-thick silver film acts as a subwavelength and deep subwavelength aperture for light and ultrasound, respectively. A spherical air bubble \cite{bubble_textbook, Lau10, Tsu14} with the $100$~nm at-rest radius is trapped and stabilised inside the hole [Fig.~\ref{fig1}(a)], which may be achieved by using a variety of established methods \cite{Tha07, Lap09, Zha13, Lau10, Tsu14, Ueh11}. In general, the bubble maintains its sphericity when it harmonically pulsates, at the microsecond scale, in response to ultrasound that evanescently enters and leaks through the hole. By solving the Rayleigh-Plesset equation of the bubble dynamics \cite{bubble_textbook, Lau10}, we demonstrate that the pulsations of the bubble allow for the tuning of light transmission as a function of the ultrasound pressure in a $200$~nm wide spectral range. This tuning mechanism opens up novel opportunities for photonics, phononics and biophysics, such as all-optical ultrasound transduction at the subwavelength scale and tuneable acousto-optic and optomechanical metamaterials.}

\begin{figure}[tb]
\centering\includegraphics[width=6cm]{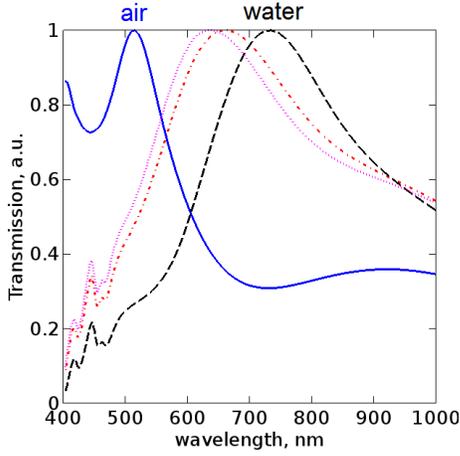}
\caption{Optical transmission through the hole. Blue solid curve: the hole is filled with air. Black dashed curve: the hole is filled with water. Magenta dotted curve: the hole is filled with water and an $R_{\rm{0}}=100$~nm air bubble. Red dashed-dotted line: the hole is filled with water and an ellipsoidal air bubble with $R_{\rm{0,x}} = R_{\rm{0,y}} =  100$~nm and $R_{\rm{0,z}} =  80$~nm at-rest radii (see Fig.~\ref{fig1} for the coordinate framework).}
\label{fig1_1}
\end{figure}

As the bubble pulsates inside the hole, the optical cutoff wavelength $\lambda_{\rm{c}}$ of the hole changes from $\sim1.7wn_{\rm{water}}$ for the entire hole filled with water to reach the asymptotic value $\sim1.7wn_{\rm{air}}$ for the air-filled hole \cite{Aba07, thesis, Mak10}, being $n_{\rm{water}}=1.33$ and $n_{\rm{air}}=1$. Thus, for $w=300$~nm the transmission will be tuned from $\sim700$~nm (at the lowest bubble radius) to $\sim510$~nm (at the largest radius), which is confirmed by $3$D finite-difference time-domain (FDTD, see \cite{Mak16}) simulations (solid and dashed curves in Fig.~\ref{fig1_1}).

For a spherical bubble in unbounded water, the pulsation implies a variation in the bubble volume, defined by the radius of the bubble $R(t)$ that harmonically varies around its at-rest radius $R_{\rm{0}}$. In both micro- \cite{Lau10} and nanobubbles \cite{Hol10, Gon10, Mao13}, $R(t)$ is described by the Rayleigh-Plesset equation and its modifications \cite{bubble_textbook, Lau10}.

Because compressibility effects are generally negligible in water \cite{Mak16}, we solve the Rayleigh-Plesset equation that models a bubble in an inviscid and incompressible liquid of constant density $\rho = 1000$~kg/m$^3$:

\begin{align}
\rho (\ddot{R} R + 1.5 \dot{R}^2) = p_{\rm{0}}\left(R_{\rm{0}}/R\right)^{3\gamma} - p_{\rm{0}} - P(t)
\label{eq:one}.
\end{align}

\noindent The initial values are $R_{\rm{0}} = 100$~nm and $\dot R = 1$~m/s \cite{bubble_textbook, Lau10}. The constant reference pressure is $p_{\rm{0}} = 100$~kPa and the polytropic exponent of air is $\gamma = 1.4$ \cite{Doi02}.

\begin{figure}[tb]
\centering\includegraphics[width=7cm]{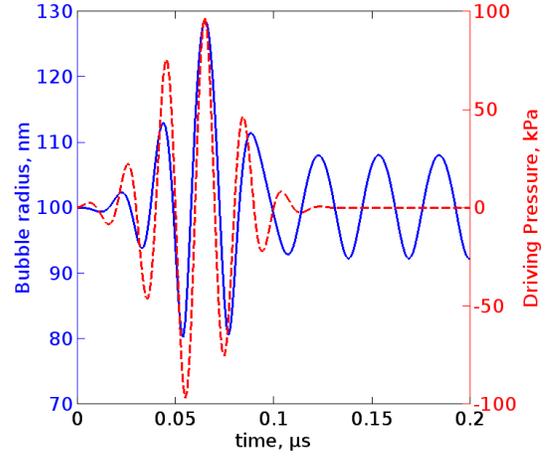}
\caption{Radial response $R(t)$ of a $100$~nm radius air bubble in water (blue solid curve) to the driving ultrasound pressure pulse $P(t)$ with the centre frequency $50$~MHz (red dashed curve).}
\label{fig2}
\end{figure}

The driving pressure $P(t)$ (red dashed curve in Fig.~\ref{fig2}) is a Gaussian-enveloped sinusoid with the frequency $f = 50$~MHz, which is detuned from the linear resonance frequency $f_{\rm{0}} \approx 30$~MHz of the bubble in water ($f_{\rm{0}}R_{\rm{0}} \approx 3$ m/s \cite{bubble_textbook, Lau10}). This detuning allows avoiding elevated acoustic forcing of the bubble to always keep $R(t)<w/2$. Because the bubble is driven off-resonance, the impact of ultrasound re-emitted due to the pulsation of the bubble is neglected. Since the bubble is securely trapped inside the hole, we also neglect the feedback between the pulsations and translational motion of the bubble \cite{Doi02, Jan09}.

Figure~\ref{fig2} (blue solid curve) shows that $R(t)$ varies in response to the driving pressure $P(t)$ with a phase lag due to the inertia of the surrounding water \cite{bubble_textbook, Lau10}. The maximum (minimum) value of $R(t)$ reached by the bubble is $\sim 130$~nm ($\sim 80$~nm). When $P(t)=0$, the bubble continues pulsating with a smaller amplitude, because no damping is considered (see \cite{bubble_textbook, Lau10}).

By using the theory from \cite{Ogu98} we show that the pulsation frequency $f_{\rm{tube}}$ of the bubble inside the rigid water-filled hole is $\sim 2$ times smaller than that in unbounded water, $f$. The numerically and experimentally verified \cite{Hyn05, Jan09, Jeu11, Qam15} theory from \cite{Ogu98} assumes that inside a rigid circular tube a pulsating bubble maintains a quasi-spherical shape, but its pulsation frequency $f_{\rm{tube}}$ decreases as

\begin{align}
\left(\frac{f_{\rm{tube}}}{f}\right)^{2} = \frac{R^{2}_{\rm{tube}}}{4R_{\rm{0}}} \left(\frac{1}{L_{\rm{1}}^{e}} + \frac{1}{L_{\rm{2}}^{e}} \right)
\label{eq:two},
\end{align}

\noindent where $L_{\rm{i}}^{e}=L_{\rm{i}}-h/2+\Delta L$ ($\rm{i}=1,2$) is the bubble position inside the tube and $R_{\rm{0}}$ is comparable with, but smaller than the radius of the tube $R_{\rm{tube}}$.  The factors $h$ and $\Delta L = 0.62R_{\rm{tube}}$ are explained in \cite{Ogu98}. Using Eq.~(\ref{eq:two}), for the considered hole we obtain $f_{\rm{tube}} \approx 0.48f$.

The wavelength of $f=50$~MHz ultrasound is $\lambda_{\rm{a}}=30$~$\mu$m (the speed of sound in water is $v_{\rm{water}} = 1500$~m/s), which is $100$ times larger than $w$. Thus, the hole operates in an acoustically deep subwavelength regime, which was not in the focus of the previous relevant works \cite{Chr07, Est08}. It is instructive to demonstrate that ultrasound may still be coupled to the hole despite the fact that the cutoff frequency $f_{\rm{c}} = 1.842 v_{\rm{water}}/(\pi w)$ \cite{boek} of the fundamental guided mode of the hole is $\sim 60$ times larger than $f$.

\begin{figure}[tb]
\centering\includegraphics[width=8.5cm]{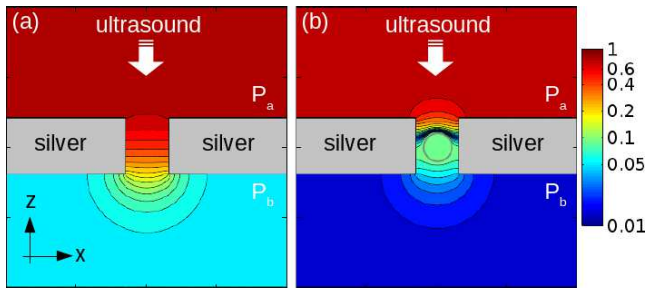}
\caption{ (a) Normalised ultrasound pressure profile (logarithmic colour bar) in the vertical cross-section of the hole without the bubble. (b) The same as in (a) but the hole is loaded with a $100$~nm radius bubble (open circle).}
\label{fig3}
\end{figure}

We simulate the two scenarios of ultrasound incident from above the water-filled hole without [Fig.~\ref{fig3}(a)] and with the air bubble [Fig.~\ref{fig3}(b)]. We use a $3$D acoustic FDTD method that models the pressure waves in water and air. However, because of a large mismatch between the characteristic specific acoustic impedance of water and that of silver, the silver film is modelled as a perfectly rigid object \cite{bubble_textbook}. The validity of this approximation was confirmed by $2$D simulations of a water-filled slit of the width $w$ in a silver film of the thickness $h$. This model takes into account the real material parameters of silver and it is less computationally demanding than a $3$D model. We revealed that ultrasound experiences strong reflections at the water-silver and silver-water interfaces of the film, which leads to a behaviour adequately reproduced by the perfectly rigid film model.  

Because $w<<\lambda_{\rm{a}}$, ultrasound is reflected from the silver film, which leads to the pressure doubling \cite{boek} in the region above the film. For clarity, in Fig.~\ref{fig3}(a) the pressure magnitude above the hole is normalised such that $P_{\rm{a}}=1$. However, ultrasound also evanescently enters the hole and decays inside it, which leads to a partial transmission of pressure $P_{\rm{b}}$ to the region below the hole. Without the bubble we obtain $P_{\rm{a}}/P_{\rm{b}} \approx 15$. In the middle of the hole, where the bubble would be located, we observe $P_{\rm{a}}/P_{\rm{bubble}} \approx 2$, which implies that a two times larger driving pressure will be required to compensate for the ultrasound attenuation in the hole. When the bubble is inside the hole [Fig.~\ref{fig3}(b)], its surface acts as a highly reflecting pressure release boundary \cite{boek}, in front of which $P \approx 0$. Nevertheless, we observe that the hole loaded with the bubble remains partially transparent, with $P_{\rm{a}}/P_{\rm{b}} \approx 50$.

\begin{figure}[tb]
\centering\includegraphics[width=8.5cm]{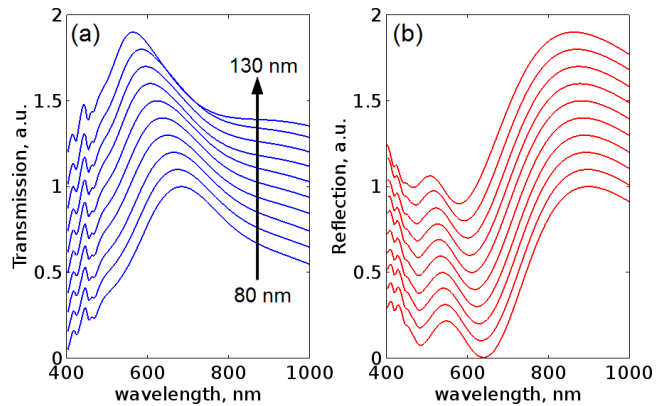}
\caption{Transmission (a) and reflection (b) spectra of the hole in the presence of the bubble at rest, plotted as a function of the bubble radius. All curves are normalised to unity and offset along the $y$-axis for the sake of visualisation. {\color{black} In Panels (a) and (b) the radius increases from $80$~nm to $130$~nm with a $5$~nm increment, as indicated by the vertical arrow in (a).}}
\label{fig4}
\end{figure}

Now we discuss the optical transmission and reflection spectra of the hole as a function of the bubble radius (Fig.~\ref{fig4}). Because the variations of $R(t)$ are much slower than the transient optical processes in the hole, the bubble is considered to be at rest but its radius takes one of the possible values of $R(t)$ in Fig.~\ref{fig2}.

In transmission [Fig.~\ref{fig4}(a)], the resonance wavelengths $\lambda_{\rm{res}}$ blueshift from $\sim 700$~nm to $\sim 525$~nm as the radius of the bubble is increased from $80$~nm to $130$~nm. The pulsations of the bubble are responsible for the blueshift, because they lead to a change in the cutoff wavelength $\lambda_{\rm{c}} \approx \lambda_{\rm{res}}$. A representative profile of the fundamental guided mode of the hole is shown in Fig.~\ref{fig1}(b). The blueshift of the minima in the reflection spectra [Fig.~\ref{fig4}(b)] is effectively smaller than in transmission, because light interacts with the hole either resonantly or nonresonantly, but interference between these two mechanisms gives rise to two differently asymmetric resonance profiles in transmission and reflection \cite{Mir10}. These asymmetries lead to a spectral offset between the corresponding transmission maxima and reflection minima.

\begin{figure}[tb]
\centering\includegraphics[width=7cm]{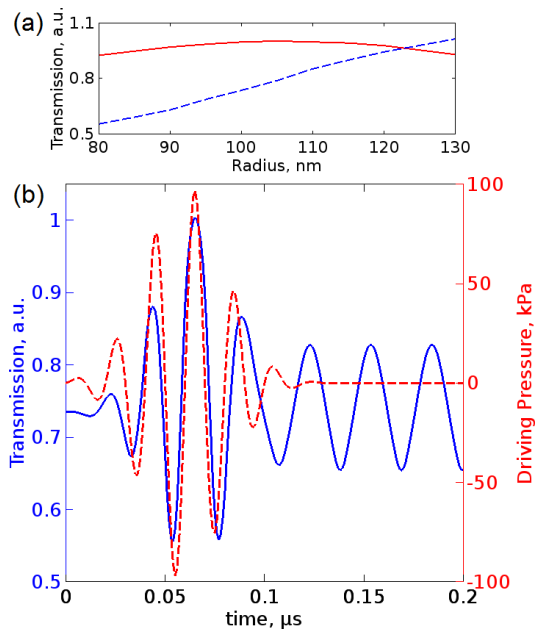}
\caption{(a) Transmitted light intensity through the hole at $561$~nm (blue dashed curved) and $632$~nm (red solid curve), plotted as a function of the bubble radius. (a) Transmitted light intensity at $561$~nm as a function of time (blue solid curve). The driving ultrasound pressure $P(t)$ (red dashed curve) can be measured by detecting the intensity.}
\label{fig6}
\end{figure}

Fig.~\ref{fig6}(a) shows the intensity of light transmitted through the hole at $561$~nm, plotted as a function of the bubble radius (blue dashed curve). The chosen wavelength corresponds to the transmission peak for the hole with the $130$~nm radius bubble. We observe a quasi-linear dependence of the intensity on the bubble radius. At $632$~nm (red solid curve), which corresponds to the peak of the light intensity transmitted through the hole with the $100$~nm radius bubble, we obtain a bell-shaped curve in which the same intensity corresponds to two different radii of the bubble. Thus, at $561$~nm each discrete light intensity can be unambiguously correlated with a unique value of $R(t)$. Also, the transmitted intensity contrast is $\sim 50\%$ at $561$~nm as compared with $\sim 15\%$ at $632$~nm. 

Fig.~\ref{fig6}(b) (blue solid curve) shows that the monitoring of the light intensity transmitted through the hole at $561$~nm allows recovering the original lineshape of $P(t)$ (red dashed curve). By fitting the the resulting curve with the Rayleigh-Plesset equation Eq.~(\ref{eq:one}) we can obtain the frequency and amplitude of $P(t)$. This functionality may be used in the optical sensing of ultrasound \cite{Mak16}.

Our discussion remains valid when the pulsating bubble becomes aspherical, which can happen, for example, because of the impact of the hole. We use simulated data for an aspherical air bubble pulsating inside a rigid tube with $\lambda_{\rm{a}}/R_{\rm{tube}}=150$ \cite{Jan09}, where $\lambda_{\rm{a}}$ is the wavelength of ultrasound. Because for our hole $\lambda_{\rm{a}}/R_{\rm{tube}}=200$, we expect a similar asphericity in our analysis. In \cite{Jan09}, the pulsating bubble was shown to be more elongated in the axial direction of the tube on expansion and in the radial direction on collapse. The maximum asphericity in the axial (radial) direction was $\sim 25\%$ ($1-2\%$). 

Transmission of light through the hole is sensitive to changes in the refractive index and effective volume of the filling material \cite{thesis}. As compared with the water-filled hole without the bubble, the maximum spectral tuning of the transmission is achieved when the hole is completely filled by an air cylinder \cite{Ili12} (solid and dashed curves in Fig.~\ref{fig1_1}). Furthermore, any intermediate air-filling shape between the sphere and the cylinder, such as an ellipsoid, gives rise to a spectral shift that is close to that produced by a spherical hole with comparable dimensions. 

A representative ellipsoidal bubble with the above discussed aspherity from \cite{Jan09} is investigated in Fig.~\ref{fig1_1} (dashed-dotted and dotted curves). We observe a small spectral change as compared with the case of the perfectly spherical bubble. Thus, it appears that bubbles of the other irregular shapes will also produce a similar spectral shift. In particular, we point out that spectral tuning will be achieved with surface nanobubbles and nanodroplets \cite{Loh15}, which may be created at the solid-liquid interface of the hole.

{\color{black} In conclusion, we have demonstrated a scheme for the control of light with ultrasound in a subwavelength hole loaded with a bubble. The scheme is expected to be used to engineer novel optomechanic devices \cite{Asp14} and acousto-optical metamaterials such as phoxonic crystals \cite{phoxonic}. For example, bubbles trapped inside the holes of the periodic photonic crystal pattern may be used to control the photonic bandgap. Bubbles may also operate inside a liquid-filled holey optical fibre \cite{Cox} or a liquid-filled dielectric slot waveguide \cite{Barrios} leading to novel opportunities for biosensing. The dependence of the bubble radius on ultrasound pressure and frequency may also be exploited in optical hydrophones \cite{Mak16}. Such devices may be used to detect bubbles in difficult to reach places such as the human brain, where gas bubbles are believed to play an important role in the Alzheimer's disease \cite{Den13}. Finally, we expect that the presented scheme can be augmented and carried out of nanophotonics back to classical microwave waveguide theory and used to solve the relevant problem of the absorptive switches \cite{IEEE}.} 

This work was supported by Australian Research Council (ARC) through its Centre of Excellence for Nanoscale BioPhotonics (CE140100003). This research was undertaken on the NCI National Facility in Canberra, Australia, which is supported by the Australian Commonwealth Government.  The authors thank B. Gibson and M. Hutchinson for useful discussions.


\begin{thebibliography}{99}

%1
\bibitem{Gen07} C. Genet and T. W. Ebbesen, Nature {\bf 445,} 39 (2007).

%2
\bibitem{Aba07} F. J. Garc\'{i}a de Abajo, Rev. Mod. Phys. {\bf 79,} 1267 (2007).

%3
\bibitem{thesis} S. Carretero Palacios, \textit{Mechanism for Enhancing the Optical Transmission through a Single Subwavelength Hole} (Prensas Universitarias de Zaragoza, Spain, 2011).

%4
\bibitem{Mak10} I. S. Maksymov, M. Besbes, J.-P. Hugonin, J. Yang, A. Beveratos, I. Sagnes, I. Robert-Philip, and P. Lalanne, Phys. Rev. Lett. {\bf 105,} 180502 (2010).

%5
\bibitem{Bul11} I. Bulu, T. Babinec, B. Hausmann, J. T Choy, and M. Loncar, Opt. Express {\bf 19,} 5268 (2011).

%6
\bibitem{Gor08} R. Gordon R, D. Sinton, K. L. Kavanagh, and A. G. Brolo, Acc. Chem. Res. {\bf 19,} 1049 (2008).

%7
\bibitem{Col09} R. M. Cole, S. Mahajan, and J. J. Baumberg, Appl. Phys. Lett. {\bf 95,} 154103 (2009).

%8
\bibitem{Liu11} Y. J. Liu, E. S. P. Leong, B. Wang, and J. H. Teng, Plasmonics {\bf 6,} 659 (2011).

%9
\bibitem{Smo02} I. I. Smolyaninov, A. V. Zayats, A. Stanishevsky, and C. C. Davis, Phys. Rev. B {\bf 66,} 205414 (2002).

%10
\bibitem{Sha07} E. A. Shaner, J. G. Cederberg, and D. Wasserman, Appl. Phys. Lett. {\bf 91,} 181110 (2007).

%11
\bibitem{Mak16_review} I. S. Maksymov, Rev. Phys. {\bf 1,} 36 (2016).

%12
\bibitem{Ger07} D. G\'{e}rard, V. Laude, B. Sadani, A. Khelif, D. Van Labeke, and B. Guizal, Phys. Rev. B {\bf 76,} 235427 (2007).

%13
\bibitem{Mak14} I. S. Maksymov and M. Kostylev, J. Appl. Phys. {\bf 101,} 084302 (2014).

%14
\bibitem{Chr07} J. Christensen, A. I. Fern\'{a}ndez-Dom\'{i}nguez, F. de Le\'{o}n-P\'{e}rez, L. Martin-Moreno, and F. J. Garc\'{i}a-Vidal, Nat. Phys. {\bf 3,} 851 (2007).

%15
\bibitem{Est08} H. Estrada, P. Candelas, A. Uris, F. Belmar, F. J. Garc\'{i}a de Abajo, and F. Meseguer, Phys. Rev. Lett. {\bf 101,} 084302 (2008).

%15_1
\bibitem{Mak15} I. S. Maksymov, Nanomaterials {\bf 5,} 577 (2015).

%15_2
\bibitem{Asp14} M. Aspelmeyer, T. J. Kippenberg, and F. Marquardt, Rev. Mod. Phys. {\bf 86,} 1391 (2014).

%15_3
\bibitem{phoxonic} A. Khelif and A. Adibi, \textit{Phononic Crystals: Fundamentals and Applications} (Springer, NY, 2016).

%16
\bibitem{bubble_textbook} T. G. Leighton, \textit{The Acoustic Bubble} (Academic Press, London, 1994).

%17
\bibitem{Lau10} W. Lauterborn and T. Kurz, Rep. Prog. Phys. {\bf 73,} 106501 (2010).

%18
\bibitem{Tsu14} H. Tsuge, \textit{Micro- and Nanobubbles: Fundamental and Applications} (Pan Stanford, Singapore, 2014).

%19
\bibitem{Lap09} D. Lapotko, Opt. Express {\bf 17,} 2538 (2009).

%20
\bibitem{Tha07} L. H. Thamdrup, F. Persson, H. Bruus, A. Kristensen, and H. Flyvbjerg, Appl. Phys. Lett. {\bf 91,} 163505 (2007).

%21
\bibitem{Ueh11} K. Uehara and Y. Yano, IEEE Trans. Magnetics \textbf{47} 2604 (2011).

%22
\bibitem{Zha13} C. Zhao, Y. Liu, Y. Zhao, N. Fang, and T. J. Huang, Nat. Commun. {\bf 4,} 2305 (2013).

%23
\bibitem{Mak16} I. S. Maksymov and A. D. Greentree, Sci. Reps. {\bf 6,} 32892 (2016).

%24
\bibitem{Hol10} R. Holyst, M. Litniewski, and P. Garstecki, Phys. Rev. E {\bf 82,} 066309 (2010).

%25
\bibitem{Gon10} M. G. Gonz\'{a}lez, X. Liu, R. Niessner, and C. Haisch, Appl. Phys. Lett. {\bf 96,} 174104 (2010).

%26
\bibitem{Mao13} Y. Mao and Y. Zhang, Nanosc. Microsc. Therm. {\bf 17,} 79 (2013).

%27
\bibitem{Doi02} A. A. Doininkov, Phys. Fluids {\bf 14,} 1420 (2002).

%28
\bibitem{Jan09} N. W. Jang, S. M. Gracewski, B. Abrahamsen, T. Buttacio, R. Halm, and D. Dalecki, J. Acoust. Soc. Am. {\bf 126,} EL34 (2009).

%29
\bibitem{Ogu98} H. N. O\~{g}uz and A. Prosperetti, J. Acoust. Soc. Am. {\bf 103,} 3301 (1998).

%30
\bibitem{Hyn05} E. Sassaroli and K. Hynynen, Phys. Med. Biol. {\bf 50,} 5293 (2005).

%31
\bibitem{Jeu11} R. Jeurissen, H. Wijshoff, M. van den Berg, H. Reinten, and D. Lohse, J. Acoust. Soc. Am. {\bf 150,} 3220 (2011).

%32
\bibitem{Qam15} A. Qamar and R. Samtaney, J. Fluid Eng. {\bf 137,} 021301 (2015).

%33
\bibitem{boek} D. T. Blackstock, \textit{Fundamentals of Physical Acoustics} (Wiley, NY, 2000).

%34
\bibitem{Mir10} A. E. Miroshnichenko, S. Flach, and Yu. S. Kivshar, Rev. Mod. Phys. {\bf 82,} 2257 (2010).

%35
\bibitem{Ili12} Y. A. Ilinskii, E. A. Zabolotskaya, T. A. Hay, and M. F. Hamilton, J. Acoust. Soc. Am. {\bf 132,} 1346 (2012).

%36
\bibitem{Loh15} D. Lohse and X. Zhang, Rev. Mod. Phys. {\bf 87,} 981 (2015).

%37
\bibitem{Cox} F. M. Cox, A. Argyros, and M. C. Large, Opt. Express {\bf 14,} 4135 (2006).

%38
\bibitem{Barrios} C. A. Barrios, Sensors {\bf 9,} 4751 (2009).

%39
\bibitem{Den13} P. A. Denis, Med. Hypotheses {\bf 81,} 976 (2013).

%40
\bibitem{IEEE} C.-H. Chen and D. Peroulis, IEEE Trans. Microwave Theory Tech. {\bf 57,} 2038 (2009).

\end{thebibliography}
\end{document}